\documentclass[]{aa}  

\usepackage[table]{xcolor}
\usepackage{txfonts,epsfig,graphicx,natbib,url,twoopt}
\usepackage{paralist}
\usepackage{booktabs,xspace}
\usepackage{amsmath}
\usepackage{tikz} 
\usepackage{amsmath}
\usepackage{amssymb}
\usepackage{graphicx}

\bibpunct{(}{)}{;}{a}{}{,}

\usepackage[hidelinks=true,breaklinks=true]{hyperref} 
\usepackage{natbib,twoopt}
\bibpunct{(}{)}{;}{a}{}{,} 
\makeatletter
 \newcommandtwoopt{\citeads}[3][][]{%
   \href{http://adsabs.harvard.edu/abs/#3}%
        {\def\hyper@linkstart##1##2{}%
         \let\hyper@linkend\@empty\citealp[#1][#2]{#3}}
	}
 \newcommandtwoopt{\citepads}[3][][]{%
   \href{http://adsabs.harvard.edu/abs/#3}%
        {\def\hyper@linkstart##1##2{}%
         \let\hyper@linkend\@empty\citep[#1][#2]{#3}}
	}
 \newcommandtwoopt{\citetads}[3][][]{%
   \href{http://adsabs.harvard.edu/abs/#3}%
        {\def\hyper@linkstart##1##2{}%
         \let\hyper@linkend\@empty\citet[#1][#2]{#3}}
	}
 \newcommandtwoopt{\citeyearads}[3][][]{%
   \href{http://adsabs.harvard.edu/abs/#3}%
        {\def\hyper@linkstart##1##2{}%
         \let\hyper@linkend\@empty\citeyear[#1][#2]{#3}}
	}
\makeatother

\renewcommand{\vec}{\boldsymbol} 

\begin{document} 
	\title{The {\em Tycho}--{\em Gaia} astrometric solution }
	\subtitle{How to get 2.5 million parallaxes with less than one year of {\em Gaia} data}
	\author{
		Daniel Michalik
			\and
		Lennart Lindegren
			\and
		David Hobbs
	}

	\institute{
		Lund Observatory, Lund University, Box 43, SE-22100 Lund, Sweden\\
		\email{[daniel.michalik; lennart; david]@astro.lu.se}
	}

	\date{Accepted for publication in A\&A, 24 Dec 2014.}

	\abstract
		{The first release of astrometric data from {\em Gaia} will contain the mean stellar positions and magnitudes
from the first year of observations, and proper motions from the combination of {\em Gaia} data with {\sc Hipparcos} prior information (HTPM).}
		{We study the potential of using the positions from the {\em
Tycho-2} Catalogue as additional information for a joint solution with early {\em
Gaia} data. We call this the {\em Tycho}--{\em Gaia} astrometric solution (TGAS).}
		{We adapt {\em Gaia}'s Astrometric Global Iterative Solution (AGIS) 
to incorporate {\em Tycho} information, and use simulated {\em Gaia} observations to 
demonstrate the feasibility of TGAS and to estimate its performance.}
		{Using six to twelve months of {\em Gaia} data, TGAS could deliver
positions, parallaxes and annual proper motions for the 2.5 million {\em
Tycho-2} stars, with sub-milliarcsecond
accuracy. TGAS overcomes some of the limitations of the HTPM project and allows its
execution half a year earlier. Furthermore, if the parallaxes from {\sc Hipparcos} 
are not incorporated in the solution, they can be used as a consistency check
of the TGAS/HTPM solution.
		 }
		{}

\maketitle
%

\section{Introduction}

The ESA astrometry satellite {\em Gaia} was launched in December 2013 with the
aim of mapping more than a billion stars ($V\la 20$) in our Galaxy
\citep{2001A&A...369..339P,2012Ap&SS.341...31D}. 
For stars brighter than $V=15$~mag, it is expected to yield positions,
parallaxes and annual proper motions at an accuracy level of 5--25~$\mu$as. 
This accuracy can only be achieved after a global reduction of observations
collected over an extended period of time (nominally five years), during which
each star is seen crossing the focal plane of {\em Gaia} on average about 70
times. The multiple observations of a given star over several years are crucial
for a successful disentanglement of the effects of stellar parallax and proper
motion. A certain redundancy of observations is also required to estimate the additional parameters for the spacecraft attitude and calibration.

While the final {\em Gaia} results are thus expected post-2020, intermediate
(provisional and less accurate) releases of astrometric data will be made; the
first one is expected in mid-2016. Being based on a much shorter stretch of
observations, it is envisaged that this first release will only give the mean
positions of the stars, as the remaining parameters may not be reliably
resolved. In previous work (\citealt{2014A&A...571A..85M}, hereafter the `HTPM
paper') we have shown that the inclusion of {\sc Hipparcos} data permits us to
compute an astrometric solution for all five astrometric parameters of the {\sc
Hipparcos} stars, based on only one year of {\em Gaia} observations. This
Hundred Thousand Proper Motions (HTPM, \citealt{LL:FM-040}) project benefits from the $\sim$24~yr
time difference between {\sc Hipparcos} and {\em Gaia} to improve the proper
motions and, for example, detect long-period astrometric binaries. However, a
serious limitation of HTPM is that the {\sc Hipparcos} stars are not numerous
enough to perform an adequate calibration and attitude determination of {\em
Gaia}. As described in the HTPM paper, additional `auxiliary stars' must
therefore be employed. Potentially this could bias the HTPM solution if the {\em Gaia} data alone do not allow all five astrometric
parameters to be determined for the auxiliary stars.

In the present paper we show that some problems with the HTPM solution can be
overcome if the auxiliary stars are replaced by stars from the {\em Tycho-2}
Catalogue \citep{2000A&A...355L..27H}, using their positions at the {\sc
Hipparcos} epoch to constrain the proper motions.\footnote{In the following, `{\em
Tycho}` always refers to the {\em Tycho-2} Catalogue.} This allows us to solve
the full set of astrometric parameters for the {\em Tycho} stars as well as for the
{\sc Hipparcos} stars, thus avoiding the potential bias from auxiliary stars.
Moreover, we find that such a solution is possible with even less {\em Gaia}
data than required for HTPM.  The resulting {\em Tycho}--{\em Gaia} Astrometric
Solution (TGAS) could become the first full-sky astrometric solution using {\em
Gaia} data, providing an important early validation of the instrument,
calibration, and data processing, at the sub-mas level.  Clearly the resulting
parallaxes and proper motions of a few million {\em Tycho} stars are extremely
interesting also from a scientific viewpoint, e.g.\ for local Galactic dynamics
and cluster studies.

{\em Tycho} refers to the star catalogues derived from the star mapper instrument
of the {\sc Hipparcos} satellite.  The original {\em Tycho-1} Catalogue
\citep{1997ESASP1200.....P} gave positions and magnitudes for about 1~million
stars. The later reduction, {\em Tycho-2} \citep{2000A&A...355L..27H}, extended
this to about 2.5~million stars, almost complete to $V\la 11.5$, and with
uncertainties of 5--70~mas at the mean epoch of observation ($\sim$1991.25).
{\em Tycho-2} also gives proper motions, derived from a comparison with old
photographic catalogues. These proper motions have uncertainties of a few
mas~yr$^{-1}$, but as they may contain systematic errors from the old data,
they are not used in the TGAS solution. 
  
\begin{figure}
\centering
\includegraphics[width=0.45\textwidth]{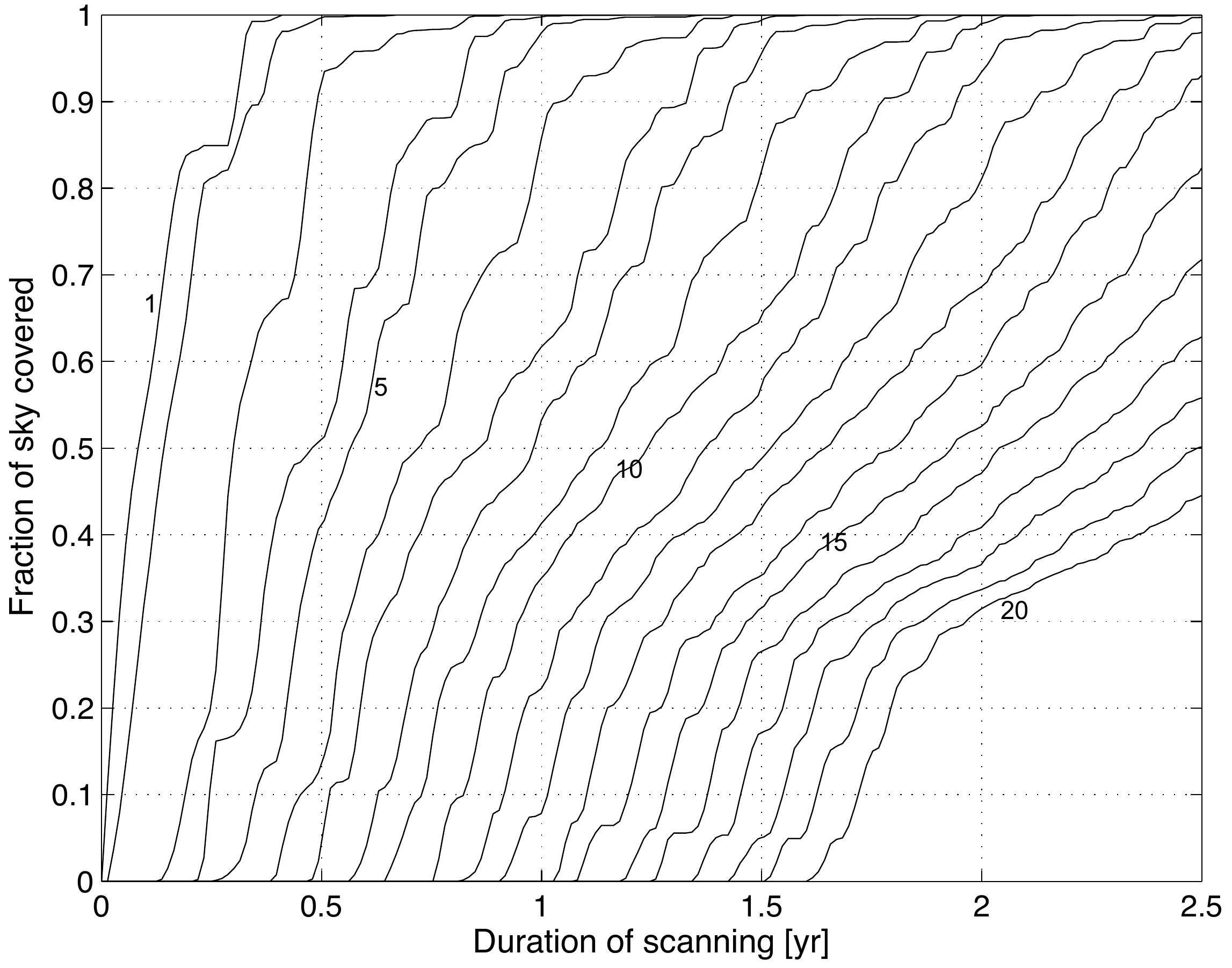}
\caption{Fractions of the celestial sphere covered by 1, 2, $\dots$, 20
distinct observations according to the nominal scanning law of {\em Gaia}, as
functions of duration. No dead time is assumed.\label{fig:frac}}
\end{figure}

\section{Prerequisites for a {\em Tycho}--{\em Gaia} solution}

\subsection{How much {\em Gaia} data are needed?}

A good astrometric solution for (apparently) single stars requires that five
astrometric parameters $(\alpha,\delta,\varpi,\mu_{\alpha*},\mu_\delta)$ are
determined for each star \citep[e.g.,][]{2012A&A...538A..78L}. A sixth
parameter ($\mu_r$) representing the radial motion (along the line of sight) is
formally required for a complete representation of the space motion. In the
present context it can be ignored, except for a limited number of nearby,
fast-moving stars with significant 
perspective acceleration for which it is assumed to be known. Thus, at least
five distinct measurements are needed for every star, where `distinct' means
that the measurements differ significantly either in time or direction. The
scanning law of {\em Gaia} causes the direction of its spin axis to change by
$4^\circ$~day$^{-1}$ \citep{2012Ap&SS.341...31D}, so that any two scans of the
same star separated by at least 5~days may count as distinct.
Figure~\ref{fig:frac} shows that after 0.5~yr, more than 90\% of the sky is
covered by at least three distinct scans, which together with the two
measurements (in $\alpha$ and $\delta$) from {\em Tycho} should in principle
suffice to determine the five astrometric parameters. 
The scans are not purely one-dimensional, but contain some across-scan
information (Sect.~\ref{sec:sim}), which is crucial for the
determination of the satellite's attitude and calibration parameters. 

The real data are affected by significant dead time,
increasing the time needed for sufficient sky coverage (Sect.~\ref{sec:deadtime}). Considering that another
half year of scanning in principle adds full redundancy to the whole sky, we
estimate that the actual amount of {\em Gaia} data required for TGAS 
corresponds to between 0.5 and 1.0~yr including dead time. 

\subsection{Incorporating the {\em Tycho} and {\sc Hipparcos} information}
\label{sec:priorconstruction}

TGAS uses the `joint solution' method described in the HTPM paper \citep{2014A&A...571A..85M}.
That is, the prior information taken from the {\sc Hipparcos} or {\em Tycho} Catalogue is cast in the form of 
normal equations $\vec{N}_\text{pri}\vec{x}=\vec{b}_\text{pri}$ for the astrometric parameters
represented by the vector $\vec{x}$. These equations are then added to the normal equations
$\vec{N}_\text{obs}\vec{x}=\vec{b}_\text{obs}$ 
derived from the {\em Gaia} observations before calculating the solution 
$\vec{\hat{x}}=(\vec{N}_\text{pri}+\vec{N}_\text{obs})^{-1}(\vec{b}_\text{pri}+\vec{b}_\text{obs})$.
The main difference compared with HTPM concerns the setting up of the prior
information for the non-{\sc Hipparcos} stars in the {\em Tycho} Catalogue,
which is described below. For the subset of {\sc Hipparcos} stars, the prior
information is taken from \citet{book:newhip} and set up exactly as for the HTPM solution (see Sect.~2.6 in the
HTPM paper). In addition to this nominal scenario, we show in Sect.~\ref{sec:noparallaxprior} that a solution can be made without the {\sc Hipparcos} parallaxes.

For a {\em Tycho}-only star the prior information in the {\em Tycho} Catalogue consists 
of the position $\alpha$, $\delta$ at the epoch J1991.25 together with its uncertainties 
$\sigma_{\alpha*}$, $\sigma_{\delta}$ and correlation coefficient $\rho$. Remaining parameters
should be treated as essentially unknown, which means that they can be set to some arbitrary values
with very large uncertainties. For the simulated solutions in Sect.~\ref{sec:sim} they are set to zero
with uncertainties $\sigma_{\varpi} = 1000$~mas, 
$\sigma_{\mu\alpha*} = \sigma_{\mu\delta} = 1000$~mas~yr$^{-1}$, and 
$\sigma_{\mu r} = \sigma_{v r} \sigma_{\varpi}/A$, where $\sigma_{v r} = 100$~km~s$^{-1}$ 
is the prior radial velocity uncertainty and $A$ the astronomical unit (HTPM paper, Eq.~17).
The prior astrometric parameters at J1991.25 are thus taken to be $(\alpha, \delta, 0, 0, 0, 0)$
with covariance  
\begin{align}
\vec{C}_\text{pri} = \begin{bmatrix}
		\sigma_{\alpha *}^2 & \rho\sigma_{\alpha*}\sigma_{\delta} & 0 & 0 & 0 & 0\\
		\rho\sigma_{\alpha*}\sigma_{\delta} & \sigma_{\delta}^2   & 0 & 0 & 0 & 0\\
		0 & 0 & \sigma_{\varpi}^2 & 0 & 0 & 0\\
		0 & 0 & 0 & \sigma_{\mu\alpha*}^2 & 0 & 0\\
		0 & 0 & 0 & 0 & \sigma_{\mu\delta}^2 & 0\\
		0 & 0 & 0 & 0 & 0 & \sigma_{\mu r}^2\\	
	    \end{bmatrix} .
\end{align} 
The prior information including the covariance is subsequently propagated to the {\em Gaia} 
reference epoch ($\sim$2015) and $\vec{N}_\text{pri}$ is calculated as the
inverse of the propagated covariance matrix. $\vec{b}_\text{pri}$ is calculated
from the difference between the prior astrometric parameters and the current
best estimate in the solution, as described in the HTPM paper, Eq.~(18).

\section{Simulations\label{sec:sim}}

In order to study the feasibility of TGAS and its potential performance we have
made numerical simulations of joint {\em Tycho}--{\em Gaia} solutions using the
AGISLab \citep{2012A&A...543A..15H} software package. AGISLab was created at
Lund Observatory to develop and test {\em Gaia} astrometric data processing strategies.
While employing the same solution algorithms as the AGIS software used to
process the real {\em Gaia} data \citep{2012A&A...538A..78L}, it runs in a much
simplified framework which also allows us to generate simulated input data (CCD
transits). The present experiments are made in a similar manner as described in
the HTPM paper, to which we refer for details. A main difference is that the
auxiliary stars in HTPM are replaced by {\em Tycho} stars, for which prior
positions are used as described in the previous section. 

Another difference is that we make the more conservative assumption 
that calibration errors contribute a constant RMS noise of 300~$\mu$as and
1000~$\mu$as per individual CCD observation, in the along-scan and across-scan
direction, respectively.
Finally, we use a (more realistic) dynamical attitude model (DAM;
\citealt{2013A&A...551A..19R}). DAM includes
a detailed modelling of the attitude perturbations caused by a number of effects such as
micro-propulsion thruster noise and micro-meteoroid hits. 
The observations are simulated using the so-called 
`astrometric attitude' \citep{2013A&A...551A..19R}, which is the physical
attitude averaged over 
the time required for a source to cross a CCD. Most of the stars
are bright which implies that observations are gated \citep{2012SPIE.8442E..1PK}
and use a shorter integration time, resulting in a less smoothed attitude.
However, we use the attitude computed for the full CCD integration time of
4.4~s, since the additional noise contribution
of shorter integration times is less than 12.7~$\mu$as, see Table~1
in \citet{2013A&A...551A..19R}, and therefore negligible in the present context. 

The real TGAS must cope with a number of complications which are ignored in the
present experiments aiming to demonstrate the basic feasibility of the concept.
The simplifying assumptions include (i) 
that there are no data gaps in the observations;
(ii) that all stars are assumed to be single,
and their motions thus consistent with the astrometric model represented by the
five (or six) astrometric parameters; 
and (iii) that it is possible to adequately calibrate the gates used to observe the bright stars.
In Sect.~\ref{sec:limits} we briefly discuss the consequences of these simplifications.

\begin{figure*}[htbp!]
\includegraphics[width=\columnwidth]{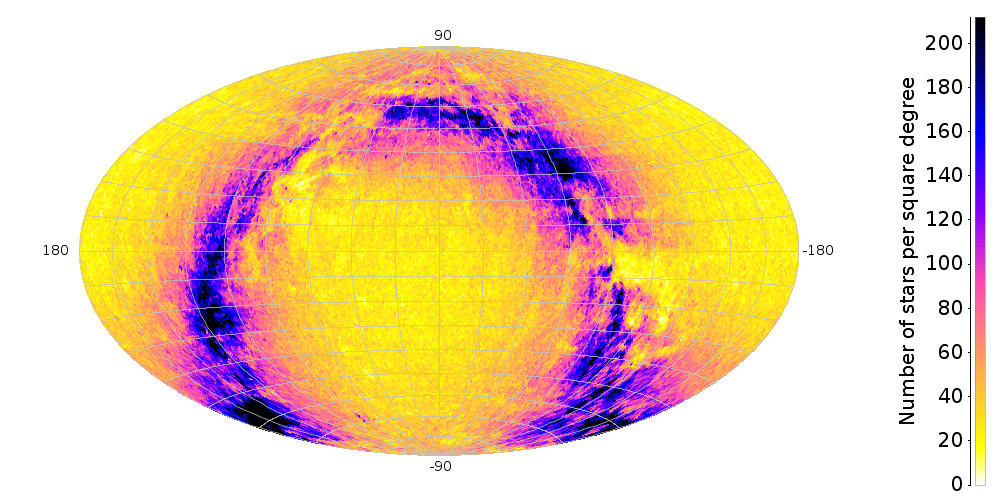}
\includegraphics[width=\columnwidth]{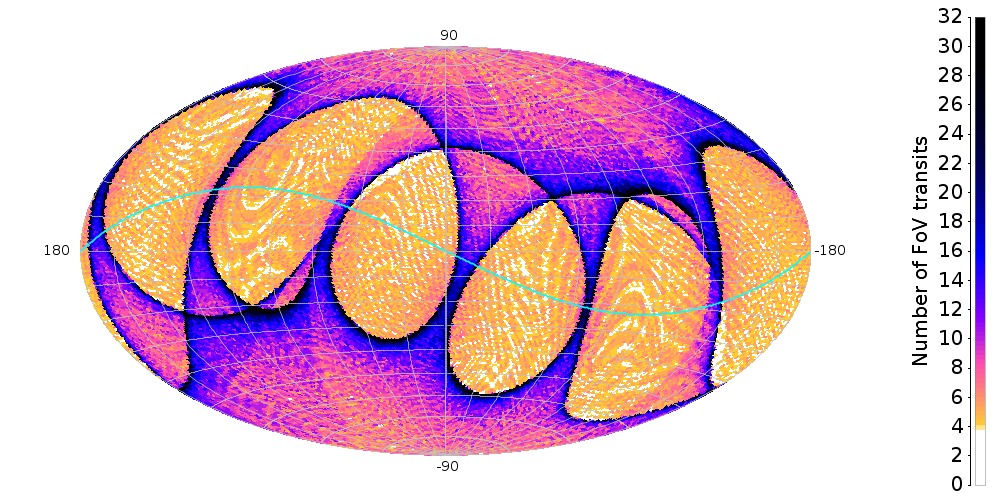}
\caption{All-sky maps in equatorial Hammer-Aitoff projection (pixel size 0.85~deg$^2$). 
{\em Left:} Stellar distribution on the sky. {\em Right:}
Number of field of view transits per star. The cyan line denotes the ecliptic.\label{fig:maps-simulations}}
\end{figure*}

The main steps of the simulations are as follows:
\begin{enumerate}
\item Astrometric parameters and uncertainties are read from the {\em Tycho} and 
{\sc Hipparcos} Catalogues and used to set up the prior information as described in
Sect.~\ref{sec:priorconstruction} and in the HTPM paper, Sects.~2.6 and 2.7. 
These parameters are also used as initial values from which the iterative astrometric 
solution is started.  
\item An artificial sky (Fig.~\ref{fig:maps-simulations}, left) is created,
representing the simulated 'true' catalogue (see 
below). This is required in order to generate {\em Gaia} observations of the stars,
and as a comparison point to evaluate the quality of the astrometric solution.
\item The {\em Gaia} observations (Fig.~\ref{fig:maps-simulations}, right) of
the {\sc Hipparcos} and {\em Tycho} stars are generated according to the
nominal scanning law, including the perturbations from DAM, and observation
noise.
For the latter we conservatively assume that all {\sc Hipparcos} and {\em
Tycho} stars are measured with the same accuracies, per CCD observation, as a
star of magnitude $G=13$, independent of the actual magnitude
($\sigma_{\textrm{AL}} = 94~\mu$as, $\sigma_{\textrm{AC}}=489~\mu$as in the
astrometric field).
\item The prior data and simulated observations are processed through the 
astrometric solution which effectively computes a least-squares estimate
of all astrometric parameters together with the parameters describing the
instrument attitude as a function of time. 
The components of the attitude quaternion are represented by cubic splines with
a knot interval of 30~s. No special provision is made to handle the rate and
angle discontinuities introduced by the use of DAM.
The astrometric solution is made for
the observation interval 2014.5--2015.0 (0.5~yr), with the reference epoch
centred on the observations, i.e., at J2014.75.
The reference frame is effectively determined by the positions and proper motions
in the {\em Hipparcos} subset.
\item The resulting astrometric catalogue is compared with the `true' catalogue
and the statistics of the differences are used to characterize the
uncertainties of the solution.
\end{enumerate}

The generation of the simulated `true' catalogue is done slightly differently for the
{\sc Hipparcos} stars and the {\em Tycho}-only stars (those that are not in the 
{\sc Hipparcos} Catalogue). For the {\sc Hipparcos} subset, `true' astrometric parameters
are simulated by perturbing the prior data (i.e., the {\sc Hipparcos} Catalogue)
by amounts that are consistent with the prior covariances. This subset 
is further described in Sect.~3.2.1 of the HTPM paper. For the {\em Tycho}-only stars,
the `true' positions are similarly obtained by perturbing the prior values according to
their assigned uncertainties and correlations. For the proper motions we regard the 
values given in {\em Tycho-2} as `true' for the present purpose; this is  
acceptable as they are not used anywhere in the solution, not even as priors.
As the {\em Tycho-2} Catalogue does not contain parallaxes, we simulate their true
values based on the apparent magnitudes, neglecting extinction and assuming that 
the absolute magnitudes have a normal distribution with mean value $+5$~mag and standard 
deviation 3~mag (see HTPM paper, footnote 4). Although it would have been possible to make the parallax
distribution dependent on the proper motion of the individual star, we do not 
consider the added complication worthwhile, as the results are rather insensitive 
to the assumed distributions. Radial velocities are simulated assuming a centred 
normal distribution with a conservatively chosen $\sigma_{v r} = 100$~km~s$^{-1}$.

\section{Results\label{sec:results}}
\begin{table}[t]
\small
\centering
\caption{Uncertainties of the astrometric parameters when processing 0.5~yr of
simulated {\em Gaia} data jointly with {\em Tycho} and {\sc Hipparcos} priors (nominal scenario).
\label{tab:results}}
\begin{tabular}{rrrrr} 
\toprule
Mag.	& Number\tablefootmark{a}	& \multicolumn{1}{c}{Position} 	
			& \multicolumn{1}{c}{Parallax}	
			& \multicolumn{1}{c}{Prop.~motion} \\
	& 	& \multicolumn{1}{c}{[$\mu$as]} 	
			& \multicolumn{1}{c}{[$\mu$as]}	
			& \multicolumn{1}{c}{[$\mu$as yr$^{-1}$]} \\
\midrule[0.2pt]                                                                   
\multicolumn{5}{c}{\cellcolor{black!10}Subset {\em Tycho} without {\sc Hipparcos}}\\
6--7	& 411		& 244 	& 399 	& 198	\\
7--8	& 8072		& 198 	& 348 	& 264	\\
8--9	& 63\,630	& 191 	& 327 	& 403	\\
9--10	& 257\,243	& 230 	& 407 	& 680	\\
10--11	& 686\,866	& 329 	& 601 	& 1145	\\
11--12	& 993\,139	& 379 	& 722 	& 1522	\\
$\ge$12	& 302\,511	& 349 	& 702 	& 1615	\\
all ($\ge$6)& 2\,311\,872		& 332 	& 631 	& 1259	\\
\midrule[0.2pt]                                                                   
\multicolumn{5}{c}{\cellcolor{black!10}Subset {\sc Hipparcos}}\\
6--7	& 9381		& 116 	& 180 	& 17	\\
7--8	& 23\,679		& 120 	& 192 	& 21	\\
8--9	& 40\,729		& 125 	& 198 	& 29	\\
9--10	& 27\,912		& 133 	& 217 	& 39	\\
10--11	& 8563		& 154 	& 253 	& 58	\\
11--12	& 2501		& 128 	& 211 	& 87	\\
$\ge$12	& 630		& 151 	& 248 	& 135	\\
all ($\ge$6)&  113\,395		& 127 	& 203 	& 32	\\
\bottomrule
\end{tabular}
\tablefoot{Nominal scenario refers to the results obtained from a simulation without data gaps
(see Sect.~\ref{sec:deadtime}) and using the full {\sc Hipparcos} prior (see
Sect.~\ref{sec:noparallaxprior}).  Uncertainties are calculated as the Robust Scatter Estimate
\citep[RSE; see footnote 18 in][]{2012A&A...538A..78L} of the differences
between estimated parameters and `true' values. 
\tablefoottext{a}{A small fraction of stars present in the {\sc Hipparcos} and
{\em Tycho} Catalogues is not observed in this simulated 0.5~yr interval of {\em Gaia}
observations.}}
\end{table} 

\subsection{Nominal scenario}
Table~\ref{tab:results} summarizes the results obtained in the nominal
scenario, i.e., using the full prior information from {\sc Hipparcos} and assuming no dead time. 
The upper part of the table gives statistics for the {\em Tycho}-only stars,
the lower part for the {\sc Hipparcos} stars. As the priors are very different
for the two subsets, they are separately discussed in the following.
\subsubsection{{\em Tycho}-only stars}

Any attempt to solve five parameters with 0.5~yr of {\em Gaia} data without a
prior utterly fails. Remarkably, however, the inclusion of the {\em Tycho}
positions allows us to solve not only the proper motions, but also the
parallaxes for the 2.5~million {\em Tycho} stars with sub-mas precision.  Here
the proper motions rely entirely on the {\em Tycho} positions, as shown by the
strong variation of the uncertainty with magnitude, mainly reflecting the
variation of positional uncertainty in the {\em Tycho} Catalogue.  In spite of
the fact that the prior parallaxes are set to zero, the posterior estimates
have very little bias (the median parallax error is $-0.7~\mu$as). 

\subsubsection{{\sc Hipparcos} stars}

It is interesting to compare the {\sc Hipparcos} subset of this solution with the
(conservative) HTPM case B, where only the positions were solved for the
auxiliary non-{\sc Hipparcos} stars (see Sect.~4.1 in the HTPM paper). The
TGAS simulation is based on half as much {\em Gaia} data as HTPM-B, uses more conservative assumptions for attitude and calibration noise, but
still provides improvements in all respects: the positions are at least a
factor two better and the proper motions improved by about 16\%.
More importantly, the resulting parallax errors are 26\% smaller and centred on zero (median error
$-0.03$~$\mu$as), while HTPM-B gave systematically underestimated 
parallaxes for the {\sc Hipparcos} stars (median error $-591~\mu$as). This
clearly demonstrates that the additional prior provided by the {\em Tycho}
positions also benefits the {\sc Hipparcos} subset.

\subsubsection{Spatial characteristics of the solution}

\begin{figure*}[htbp!]
\includegraphics[width=\columnwidth]{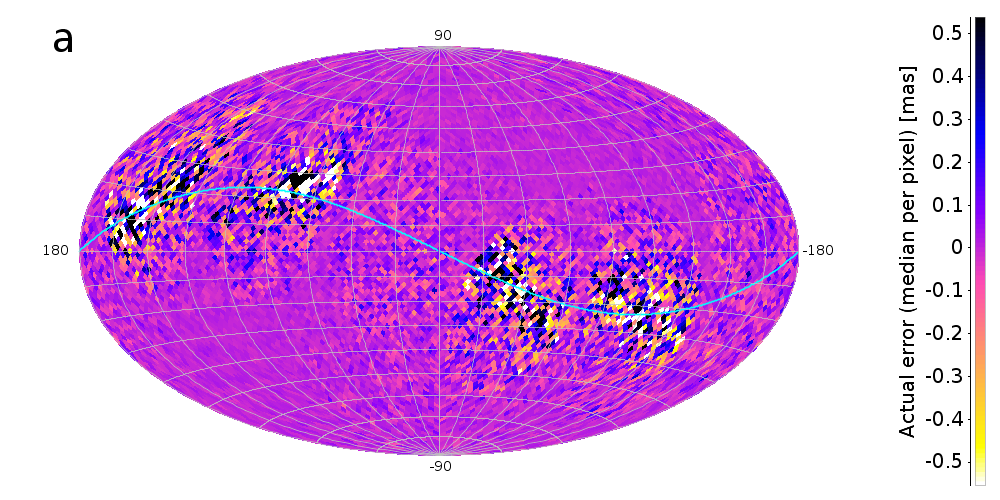}
\includegraphics[width=\columnwidth]{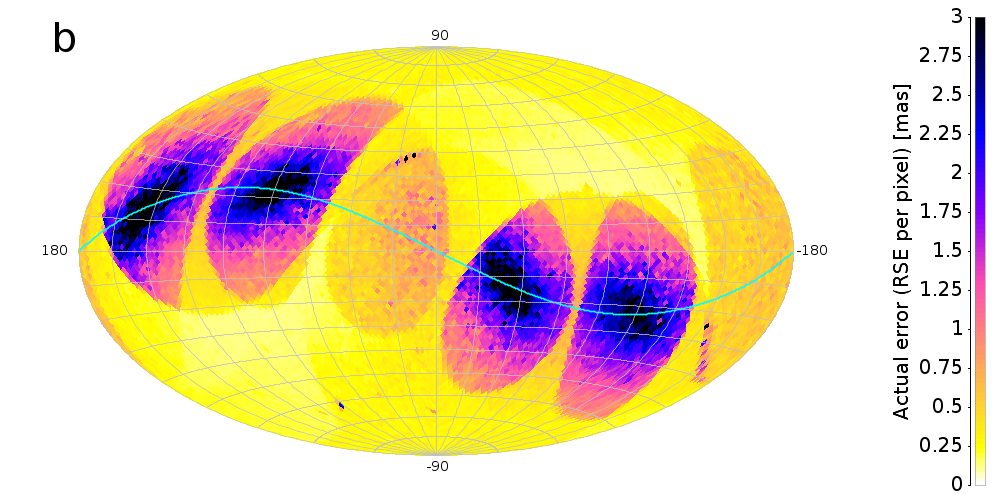}
\includegraphics[width=\columnwidth]{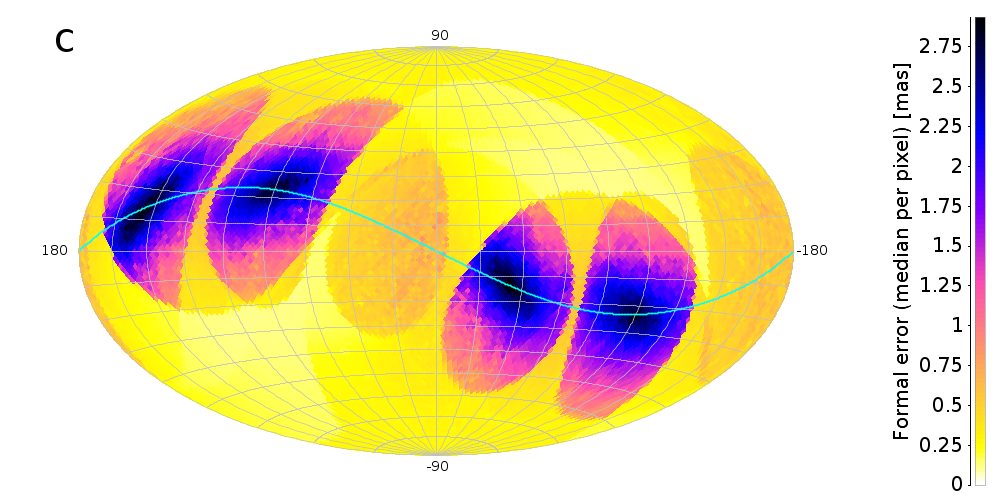}
\hfill
\includegraphics[width=\columnwidth]{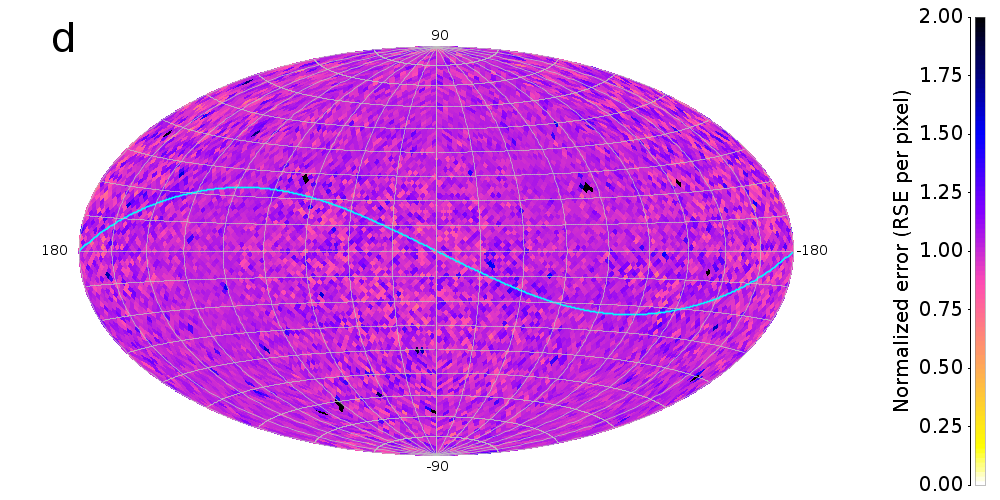}

\caption{All-sky maps characterizing the
astrometric performance of the nominal TGAS solution (pixel size
3.4~deg$^2$). The cyan line 
denotes the ecliptic. \textbf{(a)} Actual errors in
parallax (TGAS solution minus simulated true values), median per pixel to show that the solution is unbiased.
\textbf{(b)} Same as before, but RSE per pixel to characterize the size of the actual errors.
\textbf{(c)} Formal standard errors in parallax as computed in the astrometric solution.
\textbf{(d)} RSE values of the normalized errors in parallax.\label{fig:maps-results}} 
\end{figure*}

The quality of the TGAS results for a particular star depends on the number and temporal
distribution of its {\em Gaia} observations, which in turn depends on the
position in the sky. Figure~\ref{fig:maps-simulations} (right) shows the number
of field-of-view transits per star as set by the scanning law,
yielding relatively few transits in areas within 45$^\circ$ of the ecliptic.
Figure~\ref{fig:maps-results} shows the error characteristics for the {\em
Tycho}-only subset. Panel~\textbf{(a)} displays the median of the actual
parallax errors (TGAS solution minus the simulated true values). In the well
observed areas these are centred on zero, showing that the parallaxes are
unbiased. The statistical scatter is larger in areas with few observations and
unfavourable temporal distributions. There the errors could also be
correlated over several degrees. The overall median of the actual parallax
errors is $-0.6~\mu$as.  
The error maps for the other astrometric parameters have similar
characteristics. 
The size of the actual errors is shown in panel~\textbf{(b)}, displaying the RSE per pixel.

In an astrometric solution of real data the errors cannot be assessed by
comparing the solution with the true values. Error estimates must instead come from the formal
standard errors (uncertainties), computed as the square-roots of the diagonal
elements of the covariance matrix (possibly adjusted depending on the size of
the residuals in the solution). It is important that the formal standard errors
(panel~\textbf{c}) correctly characterize the actual errors.
In the ideal case, the normalized error, i.e., the ratio of the actual error to the
formal standard error, should follow a normal distribution with zero mean and unit
standard deviation all over the sky. It was already shown (by the maps of the
actual errors) that the mean values are close to zero.
Panel~\textbf{(d)} then shows the RSE values of the normalized parallax errors.
These are around 1.0 everywhere\footnote{The scale of this panel was chosen to
emphasize that most pixel values are close to 1.0. This resulted in 13 of the
12288 pixels being saturated; the three largest values are 11.2, 3.9, and 3.4.}, with a relatively small scatter in the
Galactic plane, where there are more stars per pixel. A larger scatter is seen in
the more sparsely populated areas of the sky, where the statistical uncertainty
of the calculated RSE values is higher. 
The global RSE value is 1.03, the global RMS
value 1.09. This shows that TGAS, under the given assumptions, 
provides formal standard errors that essentially correctly characterize the
actual errors.

\subsection{Simulation including data gaps\label{sec:deadtime}}
\begin{figure*}[htbp!]
\includegraphics[width=\columnwidth]{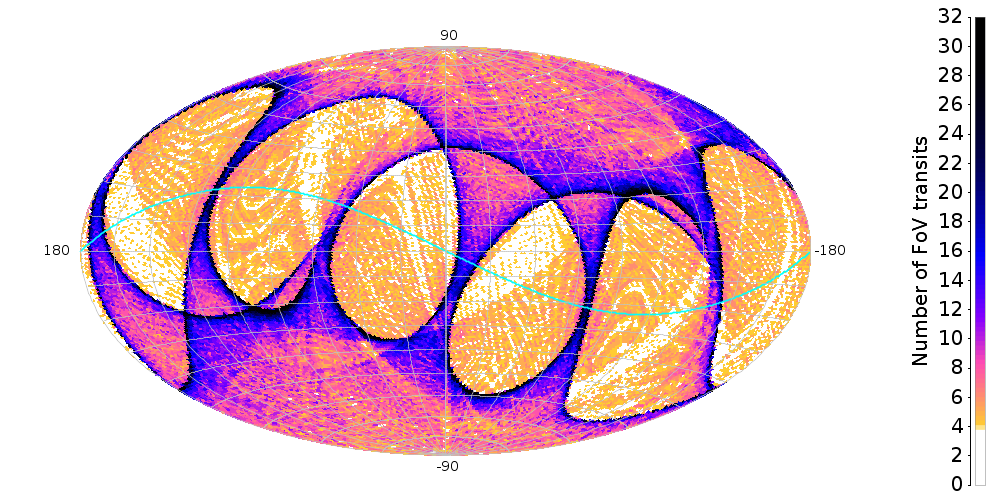}
\includegraphics[width=\columnwidth]{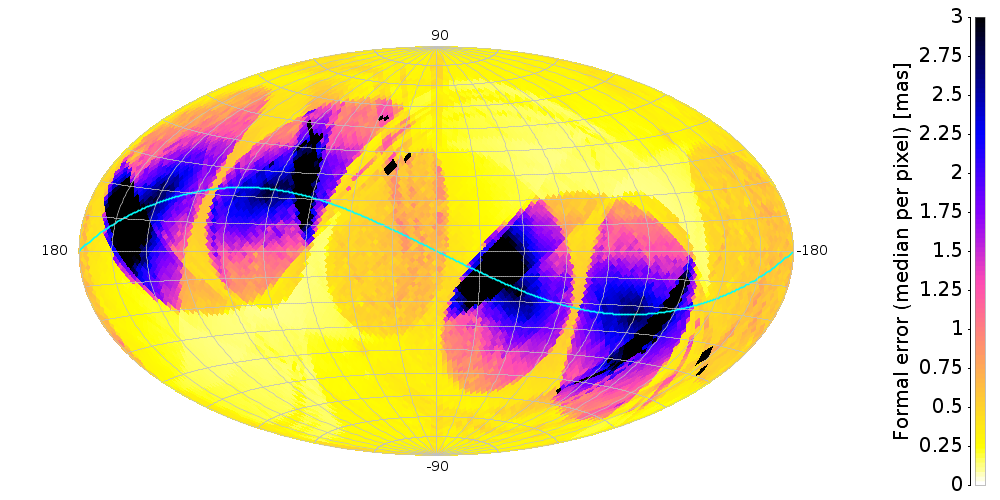}
\caption{All-sky maps for a TGAS simulation with simulated data gaps (see Sect. \ref{sec:deadtime}). {\em Left:} Number of field of view transits per star. {\em Right:} Formal standard errors (uncertainties) in parallax as computed in the astrometric solution.  \label{fig:maps-results-DT}}
\end{figure*}

The results presented so far and in Table~\ref{tab:results} are based on a simulation 
which includes all observations according to the nominal
scanning law over the assumed period of 0.5~yr. The real mission has numerous
data gaps of varying lengths, caused for example by orbit maintenance manoeuvres, 
eclipses by the moon, and solar activity. With no observations
acquired at these times, the attitude modelling cannot take advantage of the physical 
continuity of the attitude across the gaps. The result is a globally weakened astrometric 
solution, which potentially could make the TGAS solution infeasible for a dataset 
as short as 0.5~yr. We investigate this in a separate simulation including data gaps. 

The acquisition dead time for {\em Gaia} (the fraction of time during which no 
observations are acquired) is estimated to be $\simeq 6$\%. Additionally, individual 
observations may be lost due to cosmic rays, CCD defects, charge injection, telemetry 
losses, etc. Such losses are less damaging to the astrometric solution as they do not 
create gaps in the attitude determination and sky coverage, although they do affect 
the results in a statistical way. For bright stars the additional observation dead 
time is estimated to be about 5\%, resulting in a total dead time of 11\%.

To explore the robustness of the TGAS solution to acquisition dead time we apply
the actual time sequence of data gaps obtained during six months of the early {\em Gaia} 
operations\footnote{April to September 2014, including part of the commissioning phase 
which ended on July 18th.} to the nominal simulation described in Sect.~\ref{sec:sim}. 
While most of the applied gaps are shorter than 10~min, the two longest ones are 5.0 
and 2.6~days. The total length of the gaps is 15.6~days, corresponding to 8.5\% of 
acquisition dead time. Compared to the nominal value (6\%) this simulation is
therefore conservative, although we ignore the additional observation dead time.

Removing all observations corresponding to these
gaps we find that a stable solution is still possible. Compared with the solution
without gaps some stars are observed less, resulting in larger formal errors.
About 1000~stars are not observed at all, and were removed from the solution
and subsequent statistics. As shown in Fig.~\ref{fig:maps-results-DT}, the gaps
cause considerable inhomogeneity in the sky coverage and formal errors. The
white lines and wedges in the left panel show the areas most affected by the
data gaps. As expected, stars in those areas also have large formal errors, as
seen in the right panel. The formal errors are plotted on the same scale as
Fig.~\ref{fig:maps-results}, panel {\textbf c}, to show that only the areas
affected by dead time suffer from larger errors. 441 pixels are saturated, all
of these corresponding to areas affected by dead time. For 3.5\% of the sky 
the formal errors are larger than 3~mas. Globally, the parallax and position errors are
about 16\% higher than in the solution without data gaps. The proper motion
errors, which are dominated by the errors in the {\em Tycho} positions, are
less affected.

\subsection{Solution without {\sc Hipparcos} parallax prior \label{sec:noparallaxprior}}
\begin{figure*}
\includegraphics[width=\columnwidth]{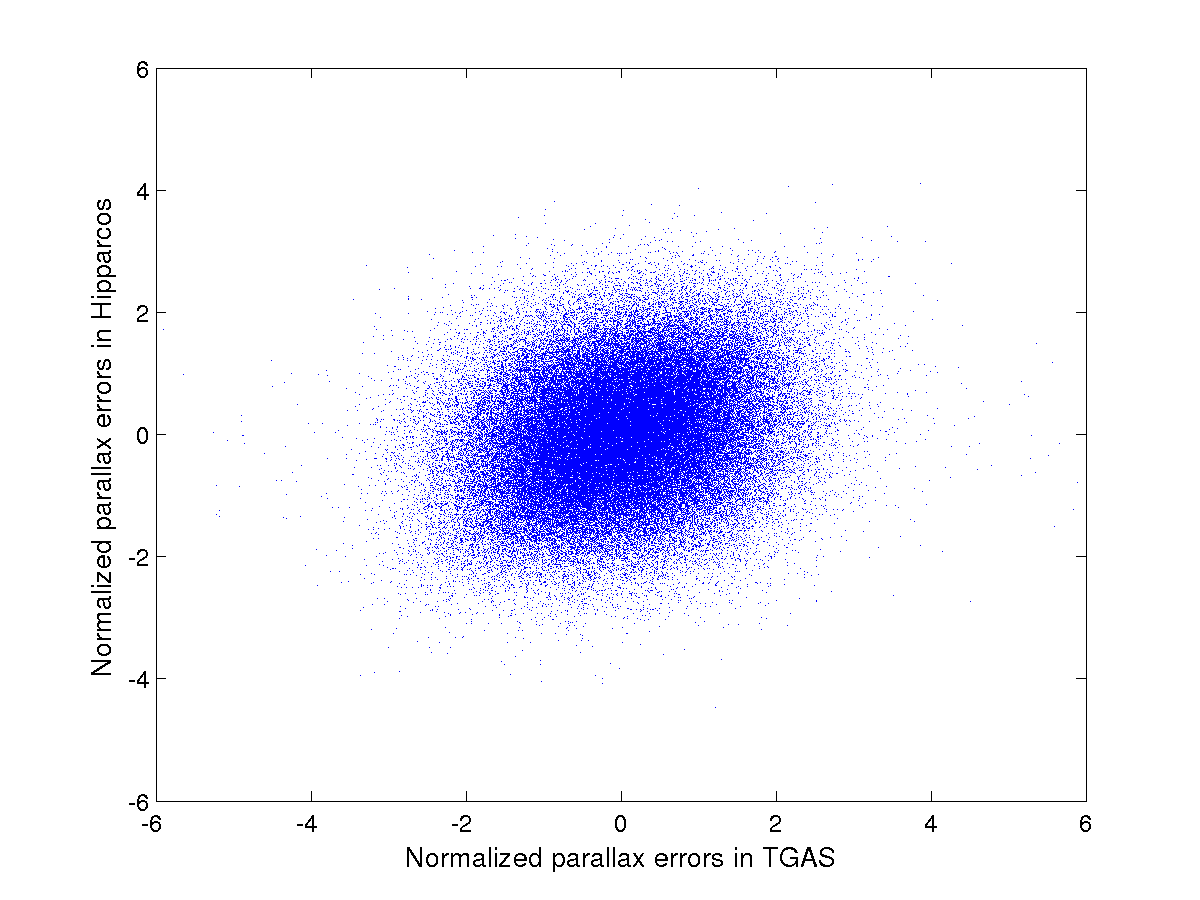}
\includegraphics[width=\columnwidth]{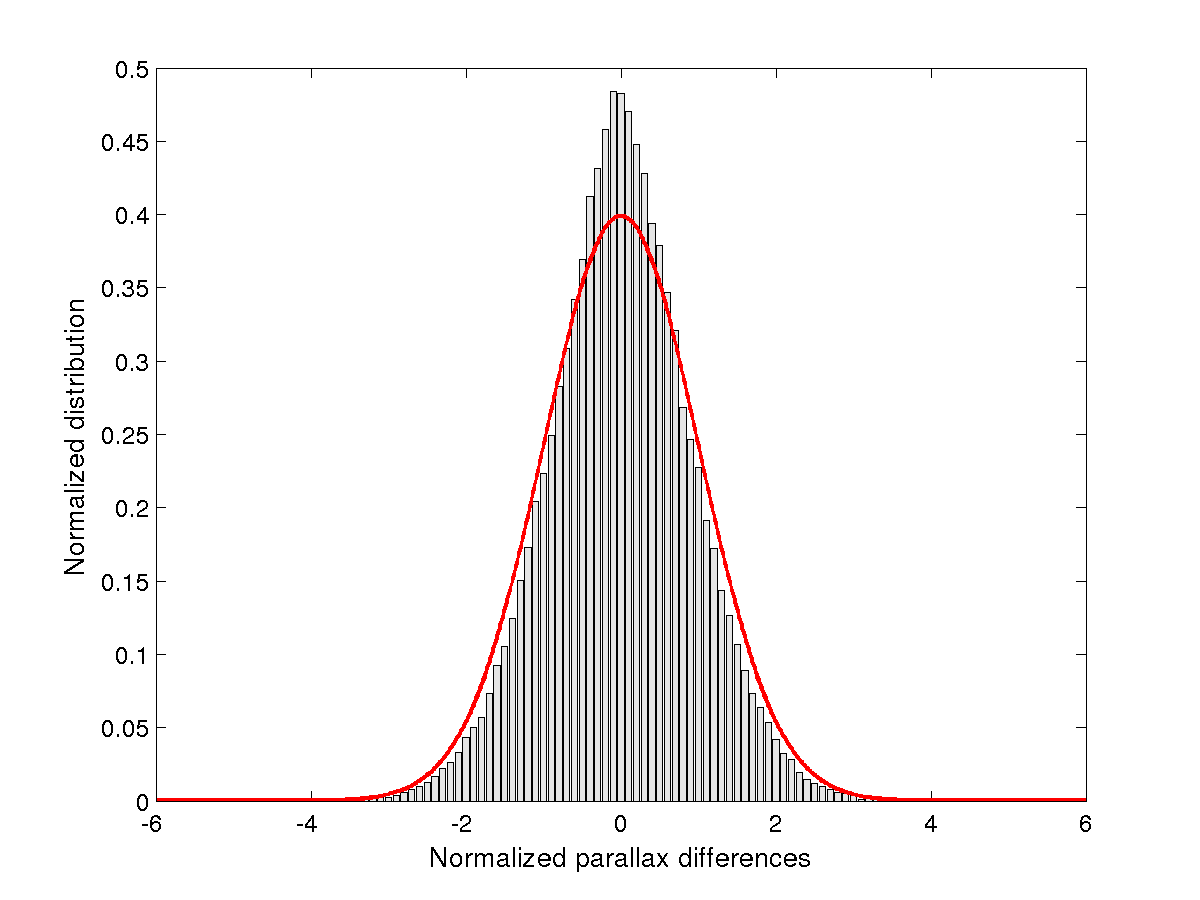}
\caption{Comparison of the parallaxes in TGAS and the {\sc Hipparcos}
Catalogue (nominal TGAS run, i.e., using the {\sc Hipparcos} parallaxes as
prior).
{\em Left:} The normalized parallax errors (calculated minus the simulated true
values, divided by their formal standard errors) are correlated. {\em Right:}
The differences of the actual parallax values (normalized by their combined
standard errors) follow a Gaussian distribution with standard deviation 0.91,
less than 1.0 because of the correlation. The solid red line is a Gaussian distribution with unit
width.\label{fig:correlated}}
\end{figure*}

For validation purposes it is desirable to compare the
parallax values of the TGAS solution with {\sc Hipparcos}. 
This is problematic when using the nominal TGAS since it already incorporates 
the {\sc Hipparcos} parallaxes as a prior. As shown in
Fig.~\ref{fig:correlated} (left panel) this leads to a statistical correlation
between the two datasets (correlation coefficient $+0.23$). As a result the
differences between the parallaxes have a smaller spread than expected from
their combined standard errors (right panel of Fig.~\ref{fig:correlated}).

\begin{figure*}
\includegraphics[width=\columnwidth]{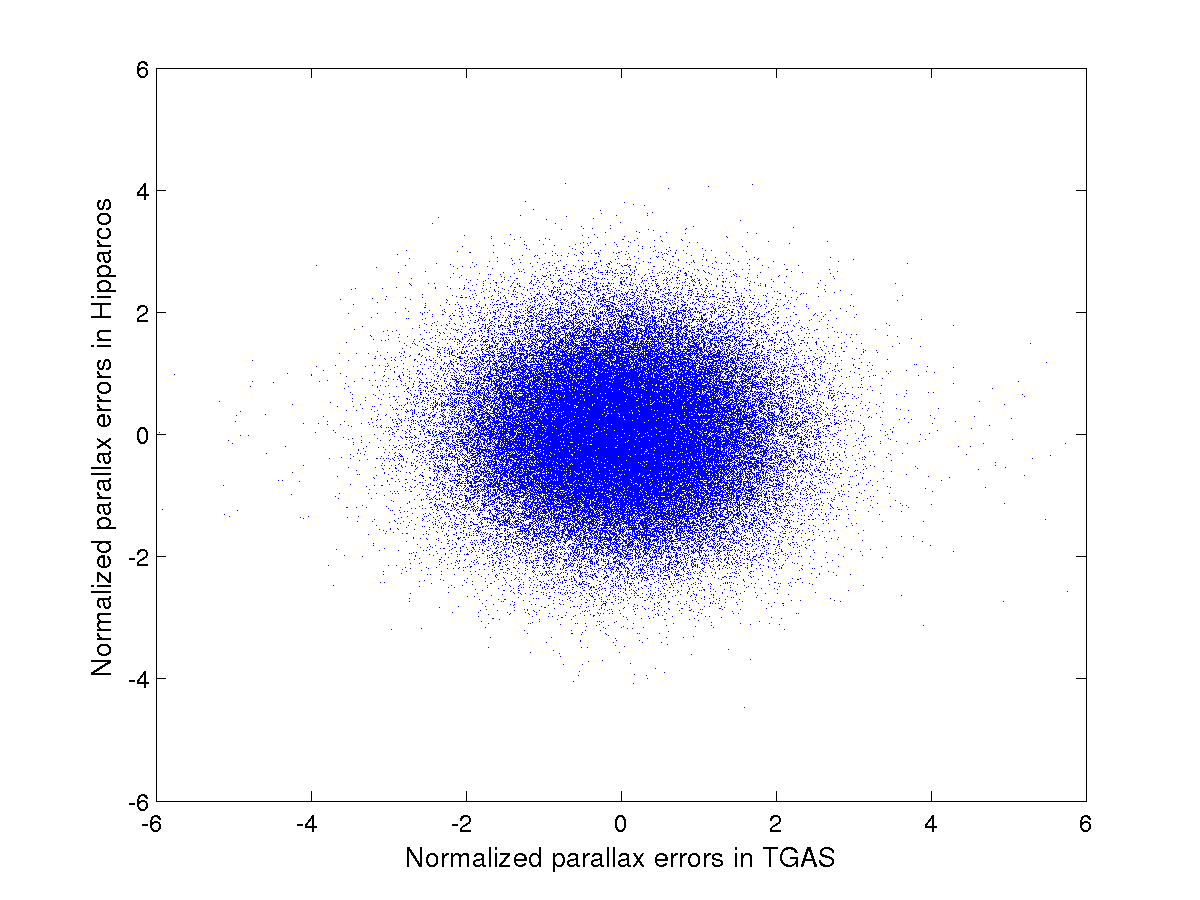}
\includegraphics[width=\columnwidth]{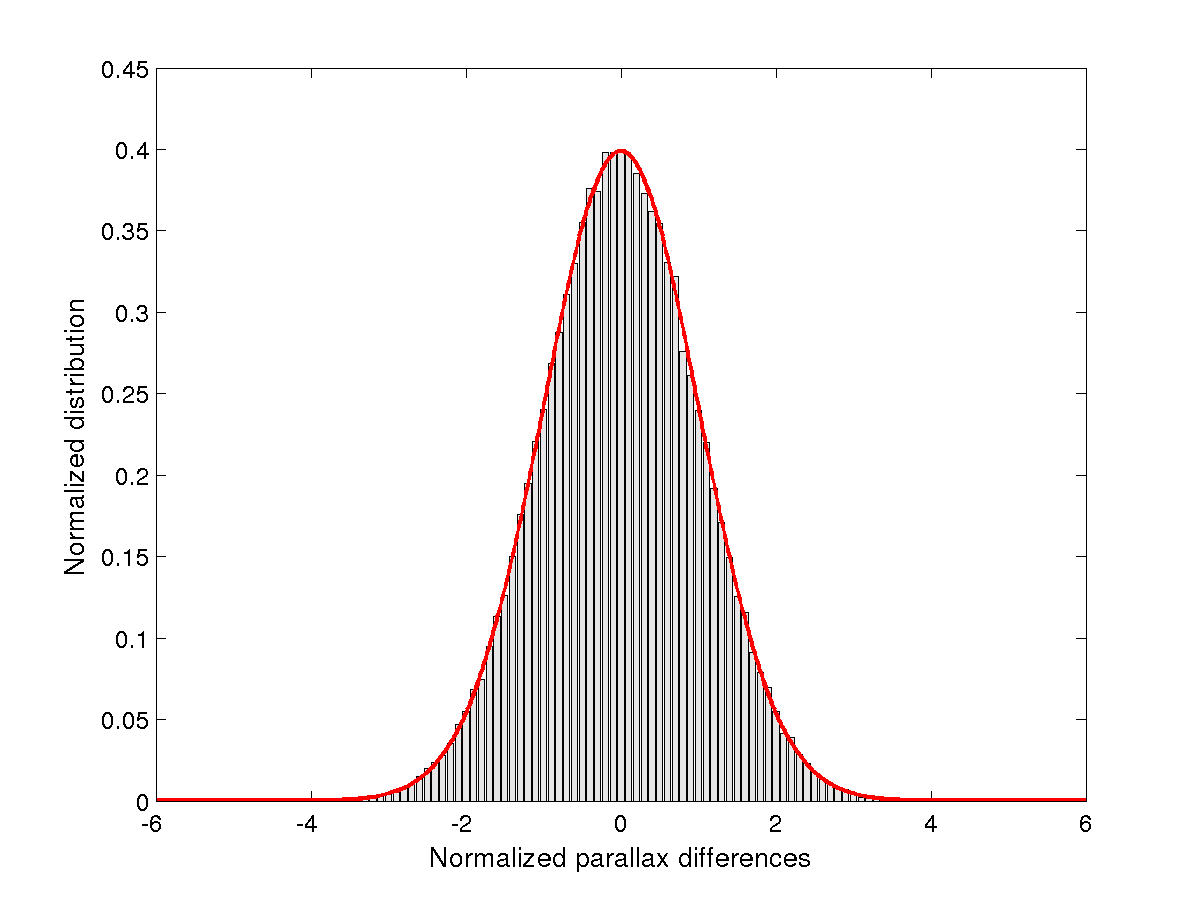}
\caption{Same as Fig.~\ref{fig:correlated}, but for the TGAS run without parallax prior (Sect.~\ref{sec:noparallaxprior}). 
{\em Left:} The normalized parallax errors are uncorrelated. {\em Right: }The
normalized differences of the actual parallax values have unit standard
deviation.
\label{fig:uncorrelated}}
\end{figure*}

To derive independent parallaxes, we propose a TGAS solution 
incorporating only the position and proper
motion information from the {\sc Hipparcos} Catalogue. This is achieved
by setting the prior parallax value and the corresponding row and column in the
{\sc Hipparcos} prior normal matrix to zero before adding the information arrays.
As shown in Fig.~\ref{fig:uncorrelated} this removes the correlation entirely
(correlation coefficient $-0.0043$) at the expense of a moderate increase in
astrometric uncertainties of the {\sc Hipparcos} subset
(Table~\ref{tab:noParallaxPrior}). The results for the {\em Tycho} subset are not shown since
the values found are virtually identical to the nominal scenario in Table~\ref{tab:results}.

\begin{table}[t]
\small
\centering
\caption{Same as Table~\ref{tab:results} (bottom), but
without incorporation of the parallax prior from {\sc
Hipparcos}.\label{tab:noParallaxPrior}}
\begin{tabular}{rrrrr} 
\toprule
Mag.	& Number	& \multicolumn{1}{c}{Position} 	
			& \multicolumn{1}{c}{Parallax}	
			& \multicolumn{1}{c}{Prop.~motion} \\
	& 	& \multicolumn{1}{c}{[$\mu$as]} 	
			& \multicolumn{1}{c}{[$\mu$as]}	
			& \multicolumn{1}{c}{[$\mu$as yr$^{-1}$]} \\
\midrule[0.2pt]                                                                   
\multicolumn{5}{c}{\cellcolor{black!10}Subset {\sc Hipparcos}}\\
6--7	& 9381		& 158 	& 270 	& 18	\\
7--8	& 23\,679	& 147 	& 241 	& 23	\\
8--9	& 40\,729	& 142 	& 232 	& 30	\\
9--10	& 27\,912	& 147 	& 244 	& 40	\\
10--11	& 8563		& 164 	& 276 	& 60	\\
11--12	& 2501		& 129 	& 212 	& 90	\\
$\ge$12	& 630		& 156 	& 251 	& 138	\\
all ($\ge$6)&  113\,395	& 147 	& 244 	& 34	\\
\bottomrule
\end{tabular}
\end{table} 

\section{Discussion}

\subsection{Consequences of the simplifying assumptions\label{sec:limits}}

\paragraph{Data gaps:}
Our simulations show that TGAS is robust to data gaps according to a  realistic
distribution of acquisition dead time. The data gaps result in an inhomogeneous
sky distribution of actual and formal errors, but do not significantly degrade
the performance in well-observed regions. Affected areas can be recovered
through additional observations after the first half year. The actual length of {\em Gaia} observations
necessary for a good solution over the whole sky is difficult to estimate without detailed
knowledge about the actual distribution of gaps. However, since the whole sky
is nominally covered by multiple scans every half year (cf.\ Fig.~\ref{fig:frac}), 
it is reasonable to conclude that the required time is less than 1~yr. 

\paragraph{Non-single stars:}
A large fraction of the TGAS stars are in reality binaries or multiple stars,
but not recognised as such in the {\em Tycho} Catalogue and thus treated as
single in TGAS. Some of them will be resolved by {\em Gaia} thanks to its higher
resolution, which makes it possible to discard these objects or treat them
appropriately. 

For systems which are unresolved also by {\em Gaia} the space motions of their
photocentres will deviate from the linear uniform model represented by the
astrometric parameters. In the later astrometric solutions of {\em Gaia} data,
objects that do not fit the single-star astrometric model will be detected and
filtered out for special treatment, but this mechanism is not effective in TGAS
due to the small redundancy of observations. TGAS will contain some fraction of
such systems with significant deviations from the adopted five-parameter model,
which will remain unrecognised in the solution. Their actual astrometric errors
will be underestimated by the formal uncertainties. This may be a common
characteristic of the early data releases, typically based on datasets with
low redundancy and imperfect calibrations. 

The impact of astrometric binaries on the derived proper motions should
nevertheless be small thanks to the 24~yr baseline of TGAS. The same is
true for the TGAS parallaxes as they are dominated by the {\em Gaia}
observations and at most only a fraction of the error in the annual proper
motion contributes to the parallax error. Comparing the proper motions from
TGAS and those in {\em Tycho-2} (which incorporate century-old ground-based
observations) could reveal not only systematic errors in the {\em Tycho-2} proper
motions but also some long-period astrometric binaries.

\paragraph{Bright-star performance and calibration issues:}
Like any other AGIS solution, TGAS will use the generic calibration model
described in \citet{2012A&A...538A..78L}, Sects.~3.4 and 3.5, 
which takes into account the actual geometry of the
optics and detectors as well as calibrations linked to chromatic image
displacements, basic-angle variations, and radiation-induced image
displacements.  
However, a specific complication of TGAS is that it almost exclusively uses stars brighter than magnitude $\sim$12, for which {\em Gaia} employs CCD gates
\citep{2012SPIE.8442E..1PK} to avoid saturation. 
The gated observations need a separate calibration
for each gate, but with the limited amount of {\em Gaia} data in a TGAS solution there may not be sufficient
observations of bright stars to do so. Gate~4 is used for the brightest stars with
magnitudes $G\la 8.84$. If it turns out that this gate cannot be reliably calibrated 
with half a year of {\em Gaia} data, we would in the worst case lose all stars brighter
than $G\simeq 8.84$ in the TGAS solution, or about 2.3\% of the {\em Tycho} stars. 
The reduced number of stars degrades the solution somewhat (for example because
the attitude is less accurately determined), but we have verified that TGAS works 
with as few as one million {\em Tycho} stars. The bright-star performance is a more 
serious issue for the HTPM solution, as more than half of the {\sc Hipparcos} stars 
are brighter than 8.84~mag. 

\subsection{Systematics in the {\em Tycho-2} data}
The present TGAS experiments assume that the {\em Tycho} positions give the
barycentric directions to the stars at the standard {\sc Hipparcos} epoch J1991.25. 
In reality the {\em Tycho-2} positions refer to slightly different epochs, which could
even be different in $\alpha$ and $\delta$. The actual TGAS solution should use the
mean effective epoch $(t_\alpha+t_\delta)/2$ of each star rather than J1991.25.

A potentially more serious complication is that the {\em Tycho} positions do not
strictly represent the barycentric directions at the given epochs of
observation. The positions were derived from the stacked star mapper photon
count records accumulated over the whole {\sc Hipparcos} mission
\citep{2000A&A...357..367H}. Parallaxes were typically not taken into
account in this process, and the resulting positions are therefore offset by a
fraction of the parallax. Both the fraction and direction of the offset depend in
a complex way on the distribution, geometry, and weights of the photon count
records. There is no simple way to correct for this effect in TGAS,
nor was it included in our simulations. However, we argue that its impact on the 
TGAS results should be very small. The
{\em Tycho} positions are mainly used to derive the proper motions on a 
baseline of 24 years. Since the parallax of a given star is typically of similar size 
as its annual proper motion, and the position offset is just a fraction of the parallax, 
it follows that the resulting annual proper motion is typically only offset by a few per 
cent of the parallax. This, in turn, should have an almost negligible impact on the 
parallax, which is mainly derived from the {\em Gaia} observations relative to
the extrapolated linear motion.

\subsection{Reference frame of TGAS}
The {\em Tycho} positions around 1991 and the {\em Gaia} observations
around 2015 are by themselves not sufficient to determine the spin of the
reference frame for TGAS, only its orientation at the {\em Tycho} epoch. 
By incorporating positions and proper motions from {\sc Hipparcos} in TGAS, in
the same way as described in the HTPM paper, the TGAS results are effectively
on the {\sc Hipparcos} reference frame.

\section{Conclusions}

The currently foreseen contents of the first {\em Gaia} data release include
positions from a two-parameter solution of the early ($\la 1$~yr) data,
because a full five-parameter solution will not be feasible, or reliable enough,
based on these data alone. Incorporating prior information into the solution 
makes it possible to solve all five astrometric parameters (i.e., including parallax 
and proper motion) with significantly less {\em Gaia} data. The HTPM project
incorporates the {\sc Hipparcos} Catalogue, resulting in greatly improved 
astrometry for the $\sim$10$^5$ {\sc Hipparcos} stars. However, as shown in 
\citet{2014A&A...571A..85M}, such a solution should be based on
at least one year of continuous {\em Gaia} data, as otherwise the results will
be biased by the use of auxiliary stars for which the full set of parameters
cannot be resolved.

TGAS extends the original HTPM proposal and takes the idea of a joint solution
one step further by combining, in a single global astrometric solution,
measurements from the early {\em Gaia} mission with data from the {\em Tycho}
and {\sc Hipparcos} Catalogues. In this paper we have shown that the 
approximate positions at the earlier epoch provided by {\em Tycho} are
sufficient to disentangle the ambiguity between parallax and proper motion 
in a short stretch of {\em Gaia} observations. Therefore TGAS allows us to 
derive positions, parallaxes, and proper motions for up to 2.5~million stars 
half a year earlier than the proposed first {\em Gaia} data release containing 
only two parameters, and one year earlier than the proposed second {\em Gaia} 
data release containing the first five parameter solution.
Using the five parameter solutions of the {\em Tycho} stars for HTPM avoids the
risk of biasing the HTPM parallaxes and improves the resulting astrometry for
the {\sc Hipparcos} stars. 
This is true even when the prior parallaxes from {\sc Hipparcos} are not used
at all in the TGAS/HTPM solution, which provides a stringent test of its consistency
with the {\sc Hipparcos} parallaxes (see Sect.~\ref{sec:noparallaxprior}). 
The moderate increase in astrometric uncertainties of such a solution compared
to the nominal scenario seems to be a price worth paying for the benefit
of a catalogue of independent parallaxes. We therefore propose that
the solution not using the {\sc Hipparcos} parallaxes should be the baseline
for TGAS/HTPM.

Our simulations of TGAS suggest that the accuracy of the resulting astrometry for
the {\em Tycho} stars will be similar to the {\sc Hipparcos} Catalogue, and
possibly significantly better depending on the exact scenario of the
number of {\em Gaia} observations available, dead time intervals, calibration,
etc. Moreover, the
dataset would be almost complete to $V\simeq 11.5$, or  3--4~magnitudes 
fainter than the survey part of the {\sc Hipparcos} Catalogue. Although the scientific lifetime of 
the data would be limited, in view of the expected later releases from 
{\em Gaia}, the potential applications cover many areas of stellar and galactic 
astronomy. Perhaps even more importantly, TGAS offers the opportunity to
perform a full-sky scientific validation of the {\em Gaia} instrument, 
calibration, and data processing at sub-mas level much earlier than previously 
anticipated. For this reason alone, we believe TGAS should be attempted as
soon as {\em Gaia} has collected sufficient data for such a solution, which could be in early
2015.


\begin{acknowledgements}
TGAS originated from discussions with Thierry Forveille and Claus Fabricius
during the review phase of the HTPM paper. We are grateful to Ulrich Bastian,
Anthony Brown, Jos de Bruijne, José Hernández, Sergei Klioner, Uwe Lammers, 
Paul McMillan, and the referee F.~van Leeuwen, for providing many supportive comments, questions, and feedback on
the manuscript. The DAM data were kindly provided by Daniel Risquez.
We gratefully acknowledge support from the Swedish National
Space Board and the Royal Physiographic Society in Lund.
\end{acknowledgements}

\bibliographystyle{aa} 
\bibliography{TGAS,agis} 

\begin{thebibliography}{11}
\expandafter\ifx\csname natexlab\endcsname\relax\def\natexlab#1{#1}\fi

\bibitem[{{de Bruijne}(2012)}]{2012Ap&SS.341...31D}
{de Bruijne}, J.~H.~J. 2012, \apss, 341, 31

\bibitem[{{H{\o}g} {et~al.}(2000{\natexlab{a}}){H{\o}g}, {Fabricius},
  {Makarov}, {Bastian}, {Schwekendiek}, {Wicenec}, {Urban}, {Corbin}, \&
  {Wycoff}}]{2000A&A...357..367H}
{H{\o}g}, E., {Fabricius}, C., {Makarov}, V.~V., {et~al.} 2000{\natexlab{a}},
  \aap, 357, 367

\bibitem[{{H{\o}g} {et~al.}(2000{\natexlab{b}}){H{\o}g}, {Fabricius},
  {Makarov}, {Urban}, {Corbin}, {Wycoff}, {Bastian}, {Schwekendiek}, \&
  {Wicenec}}]{2000A&A...355L..27H}
{H{\o}g}, E., {Fabricius}, C., {Makarov}, V.~V., {et~al.} 2000{\natexlab{b}},
  \aap, 355, L27

\bibitem[{{Holl} {et~al.}(2012){Holl}, {Lindegren}, \&
  {Hobbs}}]{2012A&A...543A..15H}
{Holl}, B., {Lindegren}, L., \& {Hobbs}, D. 2012, \aap, 543, A15

\bibitem[{{Kohley} {et~al.}(2012){Kohley}, {Gar{\'e}}, {V{\'e}tel}, {Marchais},
  \& {Chassat}}]{2012SPIE.8442E..1PK}
{Kohley}, R., {Gar{\'e}}, P., {V{\'e}tel}, C., {Marchais}, D., \& {Chassat}, F.
  2012, in Society of Photo-Optical Instrumentation Engineers (SPIE) Conference
  Series, Vol. 8442, Society of Photo-Optical Instrumentation Engineers (SPIE)
  Conference Series

\bibitem[{{Lindegren} {et~al.}(2012){Lindegren}, {Lammers}, {Hobbs},
  {O'Mullane}, {Bastian}, \& {Hern{\'a}ndez}}]{2012A&A...538A..78L}
{Lindegren}, L., {Lammers}, U., {Hobbs}, D., {et~al.} 2012, \aap, 538, A78

\bibitem[{{Michalik} {et~al.}(2014){Michalik}, {Lindegren}, {Hobbs}, \&
  {Lammers}}]{2014A&A...571A..85M}
{Michalik}, D., {Lindegren}, L., {Hobbs}, D., \& {Lammers}, U. 2014, \aap, 571,
  A85

\bibitem[{Mignard(2009)}]{LL:FM-040}
Mignard, F. 2009, {T}he {H}undred {T}housand {P}roper {M}otions {P}roject,
  {Gaia} Data Processing and Analysis Consortium (DPAC) technical note
  GAIA-C3-TN-OCA-FM-040,
  \url{http://www.cosmos.esa.int/web/gaia/public-dpac-documents}

\bibitem[{{Perryman} {et~al.}(2001){Perryman}, {de Boer}, {Gilmore}, {H{\o}g},
  {Lattanzi}, {Lindegren}, {Luri}, {Mignard}, {Pace}, \& {de
  Zeeuw}}]{2001A&A...369..339P}
{Perryman}, M.~A.~C., {de Boer}, K.~S., {Gilmore}, G., {et~al.} 2001, \aap,
  369, 339

\bibitem[{{Perryman} \& {ESA}(1997)}]{1997ESASP1200.....P}
{Perryman}, M.~A.~C. \& {ESA}, eds. 1997, ESA Special Publication, Vol. 1200,
  {The HIPPARCOS and TYCHO catalogues. Astrometric and photometric star
  catalogues derived from the ESA HIPPARCOS Space Astrometry Mission}

\bibitem[{{Risquez} {et~al.}(2013){Risquez}, {van Leeuwen}, \&
  {Brown}}]{2013A&A...551A..19R}
{Risquez}, D., {van Leeuwen}, F., \& {Brown}, A.~G.~A. 2013, \aap, 551, A19

\end{thebibliography}

\appendix
 
\end{document}